# Properties of Unshunted and Resistively Shunted Nb/AlOx-Al/Nb Josephson Junctions With Critical Current Densities from 0.1 mA/μm² to 1 mA/μm²

Sergey K. Tolpygo, *Member, IEEE,* Vladimir Bolkhovsky, Scott Zarr, T.J. Weir, Alex Wynn, Alexandra L. Day, L.M. Johnson, *Senior Member, IEEE*, and M.A. Gouker, *Senior Member, IEEE*

*(Invited Paper)*

*Abstract*—We investigated current-voltage characteristics of unshunted and externally shunted Josephson junctions (JJs) with high critical current densities, $J_c$ in order to extract their basic parameters and statistical characteristics for JJ modeling in superconducting integrated circuits as well as to assess their potential for future technology nodes. Nb/AlOx-Al/Nb junctions with diameters from 0.5 μm to 6 μm were fabricated using a fully planarized process with Mo or MoNx thin-film shunt resistors with sheet resistance $R_{sq} = 2$ Ω/sq and $R_{sq} = 6$ Ω/sq, respectively. We used our current standard MIT LL process node SFQ5ee to fabricate JJs with $J_c = 0.1$ mA/μm² and our new process node SFQ5hs (where 'hs' stands for high speed) to make JJs with $J_c = 0.2$ mA/μm² and higher current densities up to about 1 mA/μm². Using *LRC* resonance features on the *I-V* characteristics of shunted JJs, we extract the inductance associated with Mo shunt resistors of 1.4 pH/sq. The main part of this inductance, about 1.1 pH/sq, is the inductance of the 40-nm Mo resistor film, while the geometrical inductance of superconducting Nb wiring contributes the rest. We attribute this large inductance to 'kinetic' inductance arising from the complex conductivity of a thin normal-metal film in an electromagnetic field with angular frequency $\omega$, $\sigma(\omega) = \sigma_0/(1 + i\omega\tau)$, where $\sigma_0$ is the static conductivity and $\tau$ the electron scattering time. Using a resonance in a large-area unshunted high-$J_c$ junction excited by a resistively coupled small-area shunted JJ, we extract the Josephson plasma frequency and specific capacitance of high-$J_c$ junctions in 0.1 to 1 mA/μm² $J_c$ range. We also present data on $J_c$ targeting and JJ critical current spreads. We discuss the potential of using 0.2-mA/μm² JJs in VLSI Single Flux Quantum (SFQ) circuits and 0.5-mA/μm² JJs in high-density integrated circuits without shunt resistors.

This research is based upon work supported by the Office of the Director of National Intelligence (ODNI), Intelligence Advanced Research Projects Activity (IARPA), via Air Force Contract FA872105C0002. The views and conclusions contained herein are those of the authors and should not be interpreted as necessarily representing the official policies or endorsements, either expressed or implied, of the ODNI, IARPA, or the U.S. Government. The U.S. Government is authorized to reproduce and distribute reprints for Governmental purposes notwithstanding any copyright annotation thereon.

All authors are with the Lincoln Laboratory, Massachusetts Institute of Technology, Lexington, MA 02420 USA (e-mail: sergey.tolpygo@ll.mit.edu).

## I. INTRODUCTION

SIMPLE SUPERCONDUCTOR ELECTRONIC circuits hold records for the highest clock frequency [1] and the lowest energy dissipation per bit operation [2]. The main challenge for this technology has been scalability [3]. Recent developments of the fabrication technology for superconducting Single Flux Quantum (SFQ) circuits at MIT Lincoln Laboratory (MIT LL) [4],[5] have enabled very large scale integration (VLSI) of SFQ test circuits, specifically ac-biased shift registers [6] with 202 kbit and nearly $10^6$ Josephson junctions (JJs) [7], approaching the end of scaling predicted in [3] for the technology with one layer of resistively shunted junctions. The current MIT LL technology node SFQ5ee [5] utilizes Nb/AlOx-Al/Nb Josephson junctions with the Josephson critical current density, $J_c$ of 0.1 mA/μm². This process was designed to be used for energy-efficient superconducting multi-bit processors with 10 GHz to 20 GHz clock frequencies for high performance computing as the target application [8].

Increasing the integration scale of SFQ circuits above $10^6$ JJs per cm² requires reducing the area occupied by Josephson junctions and shunt resistors. This can be achieved by increasing $J_c$ above the current standard of 0.1 mA/μm² and going eventually to self-shunted high-$J_c$ junctions requiring no external shunting. A VLSI process with higher-$J_c$ junctions would also provide higher clock frequencies because the maximum clock frequency scales proportionally to the Josephson plasma frequency $f_p = (J_c/2\pi\Phi_0 C_s)^{1/2}$ [9], where $C_s$ is the junction specific capacitance and $\Phi_0 \equiv h/2e$ is the flux quantum.

It has been typically observed that reproducibility and parameter spreads of high-$J_c$ junctions is inferior to low-$J_c$ junctions, which limits the integration scale. The $J_c$ value of 0.1 mA/μm² has been viewed as a likely upper limit from the point of view of junction $J_c$ uniformity and reproducibility. However, there are applications, e.g., in digital signal processing, where achieving the highest clock frequency is a more important requirement than increasing the circuit complexity.

In order to address the need for a VLSI process with high-$J_c$ junctions for applications requiring high clock frequencies and to increase the integration scale, we have introduced a new technology node SFQ5hs, where 'hs' stands for 'high speed'. This node targets higher critical current densities $J_c =$ 0.2 mA/μm² and 0.5 mA/μm², but in all other respects is identical to the SFQ5ee node [5].

Designing SFQ circuits with complexities that would make them useful for applications and competitive with semiconductor circuits requires advanced simulation tools and accurate knowledge of the parameters of junctions and other circuit components because of narrow timing margins typical of the SFQ circuits. The properties of unshunted junctions with $J_c$ = 0.1 mA/μm² have been studied due to their use in MIT LL and ADP2 (AIST-Japan) processes, see [4],[5], [10]-[14] and references therein. The available information on the properties of junctions with $J_c$ = 0.2 mA/μm² and their parameter spreads is very limited [9],[15],[16] and almost completely absent for junctions with $J_c$ = 0.5 mA/μm² and higher. In this work we studied electrical properties of unshunted and resistively shunted junctions with $J_c$ from 0.1 mA/μm² to about 1 mA/μm² in order to extract (or refine) their basic parameters such as plasma frequency, specific capacitance, internal shunting resistance, external shunt inductance, critical current spreads, etc., for use in the junction models for circuit simulations, and to assess the potential of the high-$J_c$ junctions for future technology nodes.

## II. EXPERIMENTAL RESULTS

### A. Critical Current Density Dependence on Oxygen Exposure

A detailed description of our fully planarized fabrication process and its nodes SFQ4ee and SFQ5ee was given in [4],[5] and also reviewed in [3]. For the SFQ5hs node we keep the same minimum feature size of the critical wiring layers of 0.35 μm and the junction minimum diameter of 0.7 μm, but target higher Josephson critical current density of Nb/AlO$_x$/Al/Nb trilayers.

It is generally accepted that $J_c$ of Nb/AlO$_x$-Al/Nb junctions is mainly controlled by the oxygen exposure during aluminum oxidation $E = P_{O2} \cdot t$, where $P_{O2}$ is the oxygen partial pressure, and $t$ is the oxidation time. Many results summarized in [17]-[19] suggested the existence of a universal $J_c(E)$ proportional to $E^{-\alpha}$ dependence with the exponent changing from $\alpha \sim 0.4$ in the low-$J_c$ range to $\alpha \sim 1.6$ in the high-$J_c$ range at about 0.1 mA/μm². The latter, very steep, $J_c(E)$ dependence above 0.1 mA/μm² has been viewed as a serious challenge for a reproducible high-$J_c$ junction fabrication process.

In order to further examine the dependence of $J_c$ on $P_{O2}$ and $t$, we used dynamic oxidation of the aluminum layer and investigated several oxidation regimes: in pure oxygen; in O$_2$-Ar mixture with 3% of oxygen; and in O$_2$-He mixture with 3% of oxygen. Trilayer depositions were done in an Endura 5500 PVD cluster tool (Applied Materials, Inc.) using magnetron sputtering. Aluminum oxidation was done in a reactive chamber with a turbo-molecular pump.

The results are presented in Fig. 1. It also includes our results for the low-$J_c$ region accumulated over several years by using oxidation in pure oxygen in different chambers of the Endura cluster and in a second deposition system, CVC Connexion cluster tool. For a comparison we also show the results of the Stony Brook University group [20]-[22] obtained in 1996-2002 and of the HYPRES foundry [9],[15],[23],[24] obtained during 2003-2012 period. Superimposed are the $J_c \propto E^{-\alpha}$ dependences shown by blue dashed lines marked $E^{-0.4}$ and $E^{-1.6}$, fitting the data obtained in [17],[18] prior to 1995.

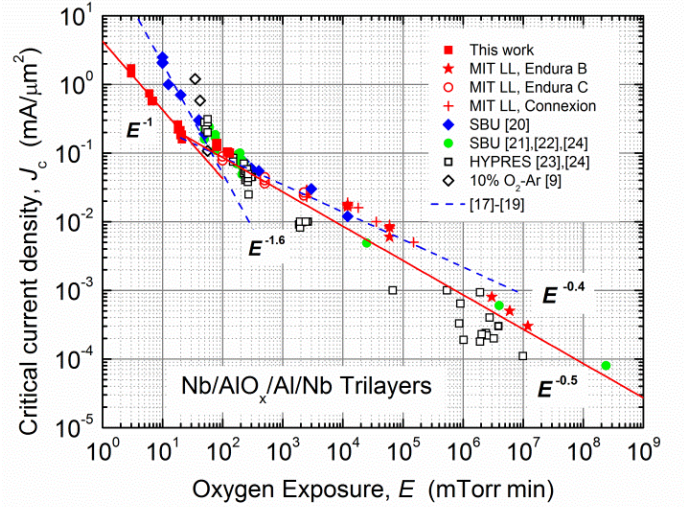

Fig. 1. Critical current density of Nb/AlO$_x$-Al/Nb junctions as a function of oxygen exposure during AlO$_x$ tunnel barrier formation by oxidation of the aluminum layer at room temperature, Historical data from [17]-[19] are shown by the blue dashed lines, corresponding to $J_c \propto E^{-1.6}$ and $E^{-0.4}$ dependences. More recent data shown by blue diamonds (♦) and green circles (●) are from the Stony Brook University group [20]-[22],[24], by open squares (□) and diamonds (◊) are from HYPRES fab [9],[23],[24]. MIT-LL data are shown by the red symbols (■), (*),(○), and (+). Our high-$J_c$ data are well described by $J_c \propto E^{-1}$ dependence (red solid line).

More detailed results for the high-$J_c$ region, obtained with the SFQ5hs process using oxidations in pure O$_2$, O$_2$-Ar, and O$_2$-He mixtures are shown in Fig. 2. We obtained $\alpha$ = 1 in the high-$J_c$ region, a much weaker $J_c(E)$ dependence than the previously reported $\alpha$ = 1.6, indicating that the exponent in $J_c \propto E^{-\alpha}$ dependence is not universal but rather depends strongly on the oxidation conditions. It also appears that the crossover into the high-$J_c$ regime in our trilayers occurs at a much lower exposure of about 20 mTorr·min and at a higher $J_c$ of about 0.2 mA/μm² rather than at 0.1 mA/μm² considered to be the boundary of the low-$J_c$ regime in the prior publications.

Comparing oxidations in pure O$_2$ and O$_2$-Ar mixtures with different dilution, we concluded that, at the same $E$, a longer oxidation at lower O$_2$ partial pressure (higher Ar dilution) results in a lower $J_c$ and higher-quality junctions than shorter oxidations at high pressures. Helium dilution resulted in the lowest $J_c$ at the same nominal oxygen exposure. It might be interesting therefore to evaluate O$_2$-Kr mixtures for barrier oxidation process.

Oxygen exposures at 20 mTorr·min and below done in pure oxygen at 4 mTorr and lower pressures did not give us reproducible $J_c$ results. However, reproducible results at $J_c$ = 0.2 mA/μm² were obtained by using at least 30-min-long oxidations in diluted O$_2$ at pressures of about 20 mTorr, thus making our SFQ5hs process with 0.2-mA/μm² JJs a stable technology node.

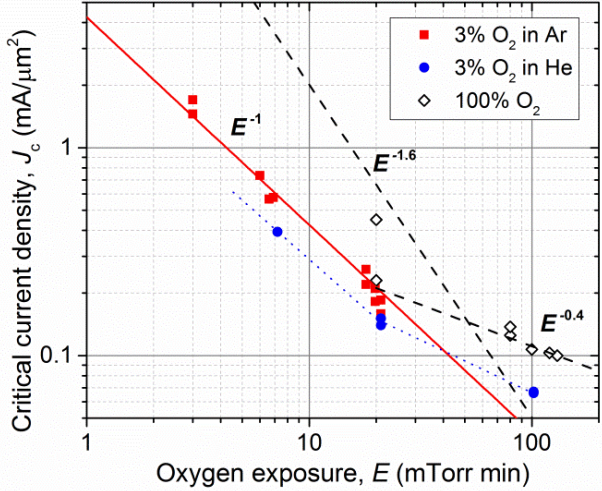

Fig. 2. High-$J_c$ region of the critical current density dependence on oxygen exposure $E = P_{O2}t$ during AlO$_x$ barrier formation in Nb/AlO$_x$-Al/Nb junctions. The solid line is $J_c \propto E^{-\alpha}$ dependence with $\alpha = 1$; the dashed lines correspond to $\alpha = 1.6$ and $\alpha = 0.4$ [17],[18]. The dotted line connecting the data points for oxidation in 3% O$_2$-He mixture is to guide the eye. The junctions were fabricated by a fully planarized process with four Nb layers. The $J_c$ was obtained from 4.2-K measurements of the critical current $I_c$ of the junctions with design diameters from 0.7 μm to 2.2 μm.

We note in this respect that $J_c \propto E^{-0.5}$ dependence typical for low-$J_c$ junctions was observed in [25] up to $J_c$ about 0.47 mA/μm$^2$, using 10-min oxidations in pure O$_2$. However, this result has not been reproduced since then. We expect that $J_c = 0.5$ mA/μm$^2$ could be also a stable technology node based on the fabrication runs produced so far. We also foresee evaluation of oxidations in 1% O$_2$ in Ar, similarly to [10].

The increase of the oxidation exponent $\alpha$ in the high-$J_c$ region is usually attributed to the appearance of defects ('pin-holes,' quantum point contacts) in the tunnel barrier, which in the superconducting state give additional channels of the subgap current transport via multiple Andreev reflections; see [26]-[28], and references therein. These defects are likely just oxygen vacancies creating atomic-size high-transparency regions in the barrier [29],[30]. We suggest that the number of these nonequilibrium defects decreases by conducting the oxidations at higher dilutions, thus maintaining a high overall gas pressure and giving oxygen more time to fill in all the required positions. The overall pressure controls the amount of collisions between the gas molecules and the wafer surface, providing momentum transfer from Ar atoms to oxygen atoms on the surface, increasing their surface mobility.

### B. I-V Characteristics of Junctions, $J_c = 0.1$ mA/μm$^2$

In order to evaluate the transport properties, we measured current-voltage ($I$-$V$) characteristics of junctions with design diameters from 0.7 μm to 2.2 μm at 4.2 K, using unshunted and resistively shunted junctions. The typical $I$-$V$ characteristics of the unshunted tunnel junctions with $J_c = 0.1$ mA/μm$^2$ fabricated by the 8-layer niobium process SFQ4ee are shown in Fig. 3.

We varied the shunt resistor in a wide range in order to achieve values of the McCumber-Stewart parameter [31],[32]

$\beta_c = 2\pi I_c R_s^2 C/\Phi_0$ from $\beta_c \sim 0.1$ to $\beta_c \gg 1$, where $R_s$ is the effective damping resistance. Fig. 4 shows the typical $I$-$V$ characteristics of the shunted junctions with $J_c = 0.1$ mA/μm$^2$ and shunts designed to target the characteristic voltage $V_c = 0.3$ mV and $\beta_c = 0.2$, where $V_c \equiv I_c R_s$.

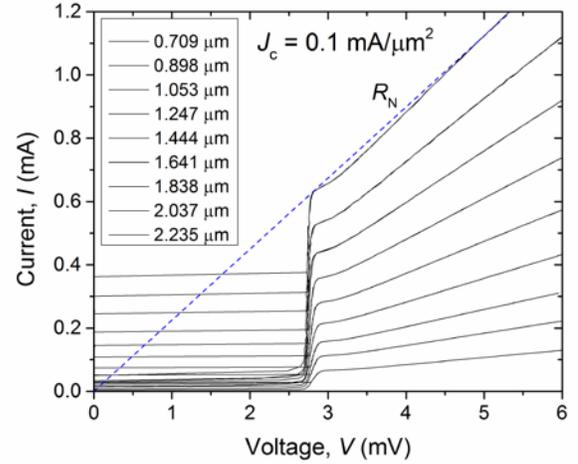

Fig. 3. The typical current-voltage characteristics of Nb/AlO$_x$-Al/Nb tunnel junctions with $J_c = 0.1$ mA/μm$^2$. Design diameter of the circular junctions used is shown in the legend, corresponding to the curves from bottom to top. The gap voltage of the junctions is 2.8 mV and $R_{sg}/R_N = 10$.

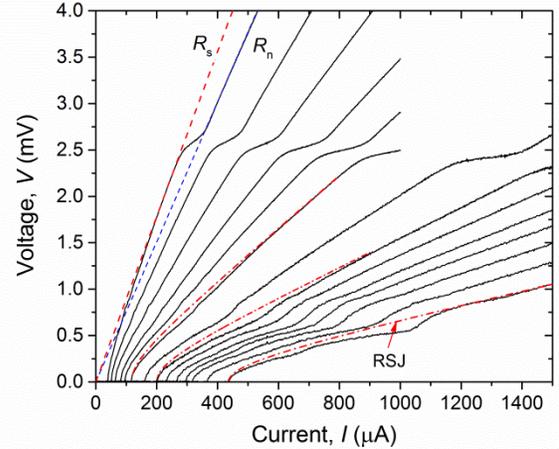

Fig. 4. Current-voltage characteristics of resistively shunted junctions at $\beta_c = 0.2$ and $J_c = 0.1$ mA/μm$^2$. The black curves from the top to bottom (and from the left to right) correspond to junctions with design diameters (in micrometers) of 0.7, 0.8, 0.9, 1.0, 1.1, 1.2, 1.4, 1.6, 1.7, 1.8, 1.9, 2.0, 2.14, and 2.34, respectively. The red and blue dashed lines are the linear approximations defining (1) and (2) in the subgap and above the gap voltage regions, respectively. Dashed-dotted red curves show the RSJ ($\beta_c \ll 1$) model dependence $V = R_s(I^2 - I_c^2)^{1/2}$ for the junctions with 1.2-μm, 1.6-μm, and 2.34-μm diameters.

In the standard junction model, it is assumed that the internal junction damping by a nonlinear (voltage-dependent) junction resistance can be replaced in the first approximation by a linear resistor $R_{sg}$ in the subgap region of voltages $V < V_g$, and by the junction normal resistance $R_N$ at voltages $V > V_g$, where $V_g = 2\Delta/e$ is the gap voltage, $\Delta$ is the superconducting energy gap in Nb electrodes. With an external shunt resistor, $R_{sh}$ in parallel with the internal resistance, the effective shunting resistance in the subgap region becomes

$$R_s = R_{sh}R_{sg}/(R_{sh} + R_{sg}), \quad V < V_g \qquad (1a)$$

and in the 'normal' resistance region

$$R_n = R_{sh}R_N/(R_{sh} + R_N), \quad V > V_g \qquad (1b)$$

The shunt resistor realized on the fabricated wafers, $R_{sh}$ may slightly differ from the design value, $R_{shd}$, resulting in $R_{sh} = kR_{shd}$ with $k \approx 1$, because the sheet resistance of the resistor films, may slightly deviate from the target values of 2 Ω/sq and 6 Ω/sq, respectively.

Fig. 5 shows the typical *I-V* characteristics of the same-size junctions with different shunt resistors. At small resistance values ($\beta_c < 0.4$), the effective shunt resistance $R_s$ can be extracted by fitting the *I-V* characteristic to the resistively shunted junction (RSJ) model $V = R_s(I^2 - I_c^2)^{1/2}$. At larger $\beta_c$ we used the slope of the *I-V* curves as shown in Figs. 4-5. Similarly, $R_n$ was obtained from the slopes of the linear dependences at $V > V_g$. The effective shunt resistances obtained in this manner are shown in Fig. 6 as a function of the design shunt resistor value for three wafers with different $R_{sg}/R_N$ ratios characterizing the tunnel barrier quality (subgap leakage). The $R_{sg}$ and $R_N$ were measured independently using unshunted junctions, and $k$ was also obtained from independent measurements of the sheet resistance and the resistor width bias.

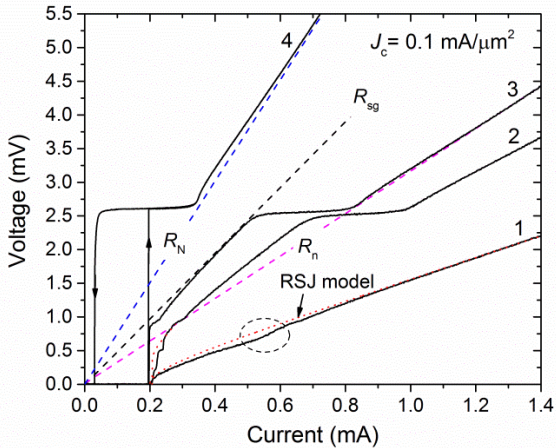

Fig. 5. Current-voltage characteristics of resistively shunted tunnel junctions of 1.6 µm diameter and $J_c = 0.1$ mA/µm². The design parameters of the shunt resistors were selected to obtain the following values of $\beta_c$ and $V_c \equiv I_cR_s$: '1' – $R_{sh} = 1.6$ Ω, $\beta_c = 0.2$, and $V_c = 0.30$ mV; '2' – $R_{sh} = 3.73$ Ω, $\beta_c = 1.0$, and $V_c = 0.69$ mV; '3' – $R_{sh} = 5.4$ Ω, $\beta_c = 2.0$, and $V_c = 0.96$ mV; '4' – unshunted junction. The dotted curves show the RSJ model ($\beta_c \ll 1$) dependence $V = R_s(I^2 - I_c^2)^{1/2}$ for '1' and resistively and capacitively shunted junction (RCSJ) model with $\beta_c = 1.0$ for '2'. Strong deviations from the RSJ model are marked in curve '1'. Note also a voltage plateau at ~ 0.5 mV in curve '2'. These features arise from a contribution of shunt inductance; see text.

As can be seen in Fig. 6, (1a) holds with a good accuracy in the entire range of resistor values studied. An excellent agreement was also obtained between the measured values of $R_n$ and (1b), giving a good self-consistency check.

Fig. 6 also stresses the importance of the subgap resistance in high-$J_c$ junctions, which can almost always be neglected in shunted low-$J_c$ junctions with $R_{sg}/R_N \gg 1$. In our $J_c = 0.1$ mA/µm² process, the typical value is $R_{sg}/R_N \approx 11$, but higher subgap leakage wafers are frequently observed with $R_{sg}/R_N$ going down to about 6.

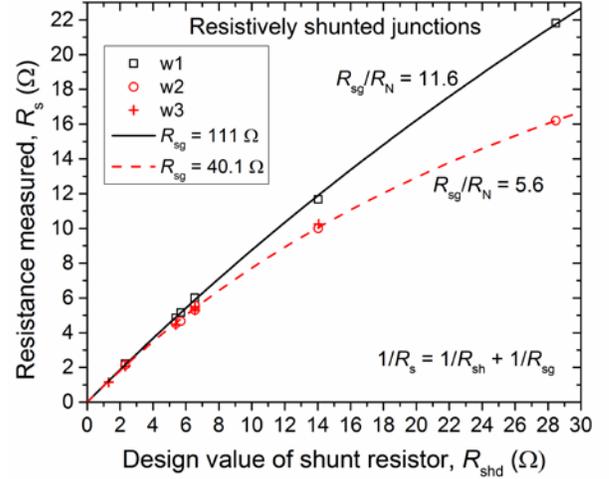

Fig. 6. Extracted values of the effective shunting resistance, $R_s$ in the subgap region as a function of design parameters of the shunt resistor, $R_{shd}$. Resistively shunted Josephson junctions with diameter of 1.6 µm were used. Solid and dashed curves show (1a) at two values of the subgap resistance, $R_{sg}$: 111 Ω, representing wafer 1 (w1), and 40.1 Ω, representing w2 and w3. The $R_{sg}$ and $R_N$ were measured independently using unshunted junctions on the same chip; $k = 0.95$ in $R_{sh} = kR_{shd}$ was determined from the sheet resistance measurements of the resistor film on the wafers at 4.2 K.

Even smaller $R_{sg}/R_N$ ratios, in the range from 3 to 6, are observed for junctions with $J_c = 0.2$ mA/µm², see Fig. 7. Subgap resistance variations may cause variability of the shunting resistance and the voltage $V_c$ characterizing the junction switching speed. A simple way to reduce this variability is to use a lower shunt resistor value, over-shunt junctions, making the relative contribution of $R_{sg}$ smaller. Over-shunting junctions reduces somewhat the maximum clock frequency but makes circuits less sensitive to the fabrication process variations.

The *I-V* characteristics of resistively shunted JJs in Fig. 5 display a few interesting features. At small $\beta_c$ values, the *I-V* curves deviate downward from the RSJ model (e.g., curve 1 in Fig. 5) at low voltages and have an upward bend and a peak in the differential resistance $dV/dI$ in the region marked in Fig. 5. At $\beta_c = 1$, the value used most often in SFQ circuit design, the *I-V* has a plateau at $V \sim 0.5$ mV and less sharp features at ~ 0.25 mV and ~ 1 mV. These features do not follow from the pure RSJ or from resistively and capacitively shunted junction (RCSJ) models but indicate the importance of the inductance associated with the resistive shunt, the so-called RLCSJ model [32]-[36], see sec. III.

### C. *I-V* Characteristics of Junctions, $J_c = 0.2$ mA/µm²

The typical *I-V* characteristics of tunnel junctions with $J_c = 0.2$ mA/µm² are shown in Fig. 7. The junctions are still highly hysteretic, although they have lower $R_{sg}/R_N$ ratios than 0.1-mA/μm² junctions, and hence require external shunting for applications in digital circuits. There is no 'excess' current; the linear part of the *I-V* characteristics at $V \gg V_g$ extrapolates to zero current, as shown by the dashed line in Fig. 7.

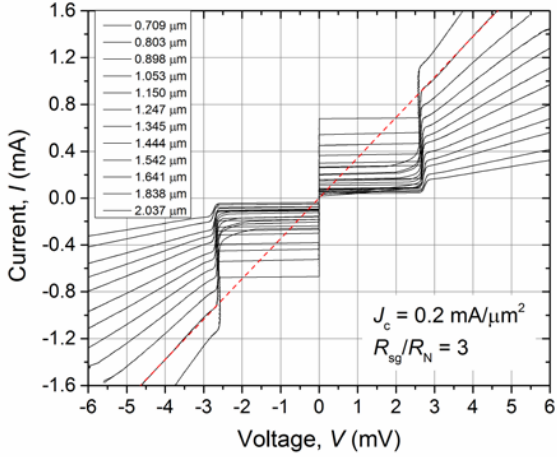

Fig. 7. Current-voltage characteristics of Nb/AlO$_x$/Nb tunnel junctions with $J_c = 0.2$ mA/μm². The curves from the top to bottom on the left side correspond to the junctions with design diameters given in the legend. The linear part of the *I-V* characteristics at $V > V_g$ extrapolates to zero as shown by the dashed line, indicating that there is no 'excess' current.

The *I-V* characteristics of the shunted junctions with 1.6-μm diameter are shown in Fig. 8. We used exactly the same design and values of the shunt resistors as for the 0.1-mA/μm² junctions. Therefore, the $\beta_c$ parameter increased proportionally to the $J_c$, and additionally because of an increase in the junction specific capacitance $C_s$, overall slightly more than by a factor of 2. Note, that, with increasing $J_c$, some features in the *I-V* curves changed significantly. For instance, a deviation from the RSJ model in Fig. 5 curve 1, became a sharp feature at the same value of $R_{sh}$, shown by a dashed circle in Fig. 8.

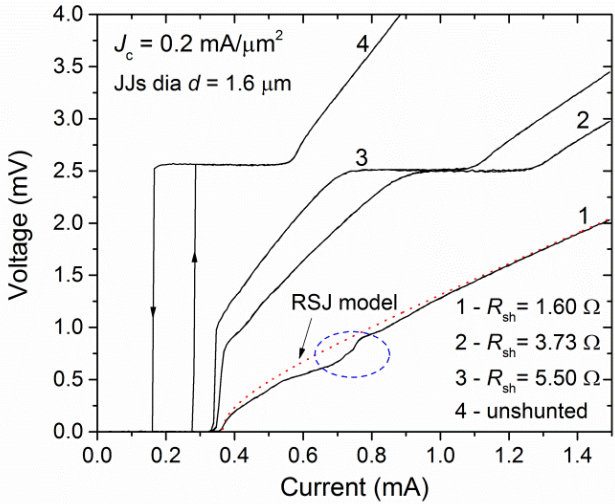

Fig. 8. Current-voltage characteristics of the resistively shunted Nb/AlO$_x$-Al/Nb tunnel junctions with $J_c = 0.2$ mA/μm². The design diameter of the junctions is 1.6 μm. The shunt resistors are exactly the same as used for 0.1-mA/μm² junctions in Fig. 5. The dotted line is the RSJ model ($\beta_c \ll 1$); the region of strong deviations from the RSJ model is circled.

Also, the dc voltage across the junction (curve 1) is lower than in the simple RSJ model in the entire range of currents. This also implies that the amplitude of SFQ pulses generated by the junction is lower than in the RSJ model.

### D. *I-V Characteristics of Junctions, $J_c = 0.8$ mA/μm²*

Fig. 9 shows *I-V* characteristics of junctions with yet higher $J_c \approx 0.77$ mA/μm². The junctions' subgap resistance is about the same as $R_N$ as shown by the dotted lines. The return (retrapping) current, $I_r$ about $0.87 I_c$. There is a large 'excess' current at $V > V_g$, i.e., the linear part of the *I-V* characteristics does not extrapolate to zero and can be described by $I = V/R_N + I_{ex}$, where $I_{ex}$ is the excess current, $I_{ex} \approx 0.36 I_c$. At large voltages the excess current diminishes, and the simple dependence $I = V/R_N$ recovers. Diminishing of the excess current is likely caused by an increase in the junction internal temperature due to Joule heating. Indeed, heating and/or nonequilibrium quasiparticle effects are indicated by a progressive decrease of the apparent gap voltage in junctions with increasing of the internal heat power $I_cV_g$. Also, a significant noise appears during *I-V* measurements at currents above about 1.6 mA, also indicating heating effects.

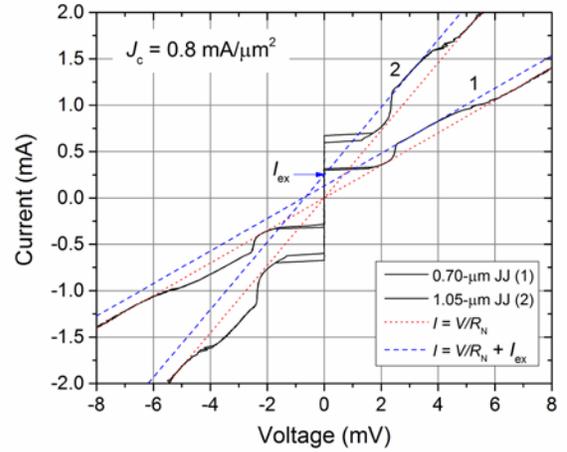

Fig. 9. The typical *I-V* characteristics of Nb/AlO$_x$-Al/Nb tunnel junctions with $J_c = 0.77$ mA/μm². The junctions become almost completely self-shunted with $I_r/I_c \approx 0.87$. Dashed lines show the presence of the 'excess' current $I_{ex}/I_c \approx 0.36$. Dotted lines show the normal-state tunneling characteristics.

It is interesting to note that the junctions switch back into the superconducting state right below a feature corresponding to a single Andreev reflection process at $V = \Delta/e$. This indicates that the subgap conductance by higher-order multiple Andreev reflection processes does not provide sufficient damping to keep the junctions nonhysteretic at this $J_c$. Nonhysteretic *I-V* characteristics were obtained at $J_c > 1$ mA/μm², not presented here.

### III. SHUNT INDUCTANCE EFFECTS

#### A. *Resistively, Inductively and Capacitively Shunted Junction*

Figs. 4-6,8 show that the *I-V* characteristics of resistively shunted high-$J_c$ junctions at large voltages are consistent with the simple linear shunt resistance approximation (1). However,

a closer inspection reveals many features of the *I-V* characteristics which do not follow from the resistively shunted junction (RSJ) model, namely voltage plateaus and current steps, which can be seen in Figs. 4,5,8. It was shown in [35]-[38] that this complex behavior is caused by the shunt inductance presumably associated with the geometrical (magnetic) inductance of the loop formed between the junction top and bottom electrodes through the resistor, as shown schematically in Fig. 10. This inductance forms a damped parallel *LRC* resonator with the junction capacitance, causing various self-induced features on *I-V* characteristics due to interaction of nonlinear Josephson oscillations with this resonator. We note in this respect a considerable recent interest in studying self-induced resonance features on *I-V* characteristics of underdamped JJs and arrays of JJs coupled to series *LC* resonators; see, e.g., [50]-[53] and references therein.

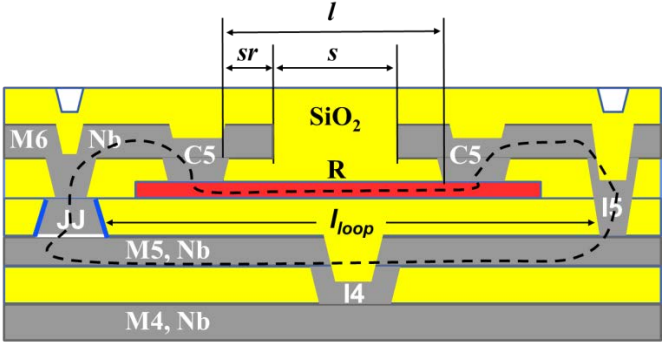

Fig. 10. A schematic cross section of a resistively shunted junction, showing a current loop (dashed black curve) associated with the shunt resistor R. The loop is formed by the junction bottom electrode M5 with width $w_{M5}$, top wire M6 with width $w_{M6}$ and the resistor. The vias to the junction, resistor, and between superconducting layers are marked C5, I5, and I4, respectively. The loop length, $l_{loop}$, is the distance between the JJ and I5 via. In all resistor designs we used $w_{M5} = w_{M6}$.

Due to inductance associated with resistive shunting, the shunted junction should be treated as resistively, inductively and capacitively shunted junction (RLCSJ) with frequency-dependent damping. Its circuit diagram is shown in Fig. 11.

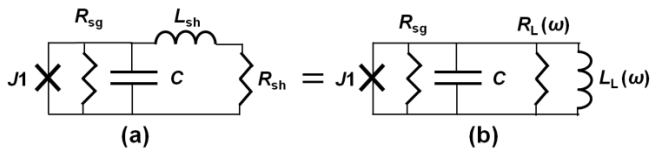

Fig. 11. (a) Shunted junction circuit diagram taking into account inductance $L_{sh}$ associated with the resistive shunt $R_{sh}$ and the junction capacitance $C$. (b) Series inductance $L_{sh}$ creates a frequency-dependent damping $R_L(\omega)$ and inductive load $L_L(\omega)$.

The junction $J1$ with its subgap resistance $R_{sg}$ can be viewed as an rf current source with a finite internal resistance. The impedance seen from this source is

$$Z(\omega) = \frac{R_{sh} + i\omega L_{sh}}{1 - \omega^2 L_{sh} C + i\omega R_{sh} C} \quad (2)$$

Its imaginary part becomes zero at a resonance frequency [33]

$$\omega_r^2 = \frac{1}{L_{sh}C} - (\frac{R_{sh}}{L_{sh}})^2 = \omega_0^2 - \omega_L^2, \quad (3)$$

where $\omega_0^2 = (L_{sh}C)^{-1}$ is the *L-C* resonance frequency in the absence of dissipation and $\omega_L = R_{sh}/L_{sh}$.

The resonance exists if $\omega_0 \geq \omega_L$, that requires $(L_{sh}/C)^{1/2} \geq R_{sh}$, i.e., the resonator impedance be larger than $R_{sh}$. This is equivalent to the requirement that the resonator be underdamped at frequency $\omega_L$, i.e., $\omega_L \tau_c \leq 1$, where $\tau_c = R_{sh}C$. The corresponding resonance voltage is given by the Josephson relation

$$V_r = \omega_r \Phi_0/(2\pi) . \quad (4)$$

At $\omega = \omega_r$, the impedance seen from the junction is real

$$Z(\omega_r)/R_{sh} = (\omega_L \tau_c)^{-1} \quad (5)$$

and larger than $R_{sh}$ because $\omega_L \tau_c < 1$. Therefore, an increase in the slope of the *I-V* characteristic, $dV/dI$ occurs near the resonance voltage.

However, the real part of impedance (2) reaches its maximum (has a 'resonance') at a higher frequency (voltage), given by

$$\omega_m^2 = \frac{1}{L_{sh}C} - \frac{1}{2}(\frac{R_{sh}}{L_{sh}})^2 = \omega_0^2 - \frac{1}{2}\omega_L^2, \quad (6a)$$

$$V_m = \omega_m \Phi_0/(2\pi), \quad (6b)$$

independently of whether $\omega_r$ exists or not. The maximum value of the real part of the impedance is

$$ReZ(\omega_m)/R_{sh} = [(\omega_L \tau_c) - \frac{1}{4}(\omega_L \tau_c)^2]^{-1} . \quad (7)$$

If we normalize voltages to $I_c R_{sh}$ product, $v = V/(I_c R_{sh})$, and introduce dimensionless shunt parameters $\beta_L = 2\pi I_c L_{sh}/\Phi_0$ and $\beta_C = 2\pi I_c R_{sh}^2 C/\Phi_0$, the resonance frequencies can be expressed as

$$v_r = \beta_L^{-1}(\frac{\beta_L}{\beta_C} - 1)^{\frac{1}{2}} \quad (8a)$$

$$v_m = \beta_L^{-1}(\frac{\beta_L}{\beta_C} - 1/2)^{1/2} , \quad (8b)$$

and the resonance existence condition as $\beta_L \geq \beta_C$.

### B. Extraction of Shunt Inductance, $L_{sh}$

A good description of all complex features in the *I-V* characteristics and the value of $L_{sh}$ can be obtained from fitting the characteristics to the RLCSJ model [36]-[38]. For SFQ circuit design it is preferable to not have internal shunt resonances as they complicate the junction dynamics. Therefore, below we will estimate the relevant parameters and the effects of the shunt design based on the position of peaks in $dV/dI$ versus $V$ dependences, leaving a more complete treatment for a separate publication.

The *I-V* characteristics of 0.1-mA/$\mu$m² JJs shown in Fig. 4 correspond to $\beta_C \approx 0.22$ and $I_c R_{sh} \approx 0.33$ mV. They all display small and relatively broad peaks in $R_{sh}^{-1}dV/dI$ in the following ranges of voltages: $0.2 - 0.25$ mV; $0.45 - 0.54$ mV; $0.64 - 0.67$ mV; $0.74 - 0.79$ mV; $0.85 - 0.95$ mV. The voltage of the strongest peak, which occurs at $V_m$, is given in Table I. For all JJ sizes, the amplitude of the strongest peak in $R_{sh}^{-1} dV/dI$ is in the range from 1.4 to 2, increasing with the junction size.

The second strongest peak is in the range from 0.2 mV to 0.25 mV, corresponding to the third subharmonic of $V_r \sim 0.67$ mV. The subharmonic features appear because at voltages $V < V_c$ the Josephson oscillations are rich in higher harmonics of the fundamental Josephson frequency $f_J = V/\Phi_0$, so the resonance at $V_r$ can be pumped at the subharmonics of $V_r$: $V_r/2$, $V_r/3$, etc. At this shunting, however, the second subharmonic feature is small because $V_r/3 < V_c < V_r/2$ and Josephson oscillations at $V > V_c$ are almost sinusoidal.

TABLE I
EXTRACTED INDUCTANCE OF MO SHUNT RESISTORS WITH $R_{sq} = 2$ Ω/SQ

| JJ diameter (μm) | Shunt resistor (Ω) | Resistor width, $w$ (μm) | $dV/dI$ peak voltage (mV) | $L_{sh}^+$ (pH) | $L_{sh}^-$ (pH) |
|---|---|---|---|---|---|
| 0.7 | 9.7 | 1 | 0.762 | 6.60 | 1.33 |
| 0.8 | 7.1 | 1 | 0.744 | 5.21 | 0.95 |
| 0.9 | 5.5 | 1 | 0.790 | 3.46 | 0.76 |
| 1.0 | 4.3 | 1 | 0.752 | 3.15 | 0.56 |
| 1.1 | 3.5 | 1 | 0.798 | 2.23 | 0.47 |
| 1.2 | 2.9 | 1 | 0.798 | 1.86 | 0.38 |
| 1.4 | 2.08 | 1.2 | 0.798 | 1.36 | 0.27 |
| 1.6 | 1.6 | 1.5 | 0.772 | 1.11 | 0.21 |
| 1.7 | 1.4 | 2.0 | 0.760 | 1.03 | 0.18 |
| 1.8 | 1.25 | 2.0 | 0.744 | 0.960 | 0.16 |
| 1.9 | 1.13 | 2.0 | 0.722 | 0.920 | 0.14 |
| 2.0 | 1.0 | 2.1 | 0.710 | 0.868 | 0.12 |
| 2.335 | 0.733 | 3 | 0.660 | 0.749 | 0.09 |

The shunt inductance can be estimated from (6) using the voltage corresponding to the main peak in $dV/dI$ and capacitance $C = C_s A_J$, where $A_J$ is the junction actual area [4] and $C_s$ the specific capacitance. For 0.1-mA/μm$^2$ junctions, we used $C_s = 70$ fF/μm$^2$, see Sec. IV and [40]. Both equations (3) and (6) are quadratic and have two solutions for $L_{sh}$, i.e., the same resonance frequency can be obtained using two very different inductors – a unique feature of $LRC$ resonators. Table I summarizes parameters of molybdenum resistors with 2 Ω/sq sheet resistance used to shunt the JJs in Fig. 4, the voltage of the $dV/dI$ main peak, and the two solutions for the shunt inductance $L_{sh}^+$ and $L_{sh}^-$ obtained.

$L_{sh}^-$ is an extraneous solution because, in all cases, it corresponds to a resonator with quality factor $Q < 1$, where $Q = \sqrt{\frac{L_{sh}}{C}}/R_{sh}$. This resonance would not be observable. On the other hand, $L_{sh}^+$ gives $Q$ in the range from 1.47 to 2, which is consistent with the observed $R_{sh}^{-1} dV/dI$ peak amplitudes in the range from 1.44 to 2. Therefore, we will use hereafter $L_{sh}^+$ for the total shunt inductance.

As an additional verification, we simulated the $I$-$V$ characteristics of a shunted JJs using $L_{sh}^+$ in series with the shunt resistor and compared it with the measurements in Fig. 12a. We see that the simulated $I$-$V$ agrees very well with the measured one and captures all main features of the measured characteristics: overall reduction in the time-averaged voltage with respect to the RSJ model; voltage upturn (increase in $dV/dI$) at $V \approx V_r$; and the change in curvature of $V(I)$, i.e., a maximum in $dV/dI$, at $V \approx V_m$.

Fig. 12b shows the simulated voltage waveforms at a current slightly above the $I_c$ in the case of a purely resistive shunt (dashed-dotted curve) and with the shunt inductance taken into account (solid curve).

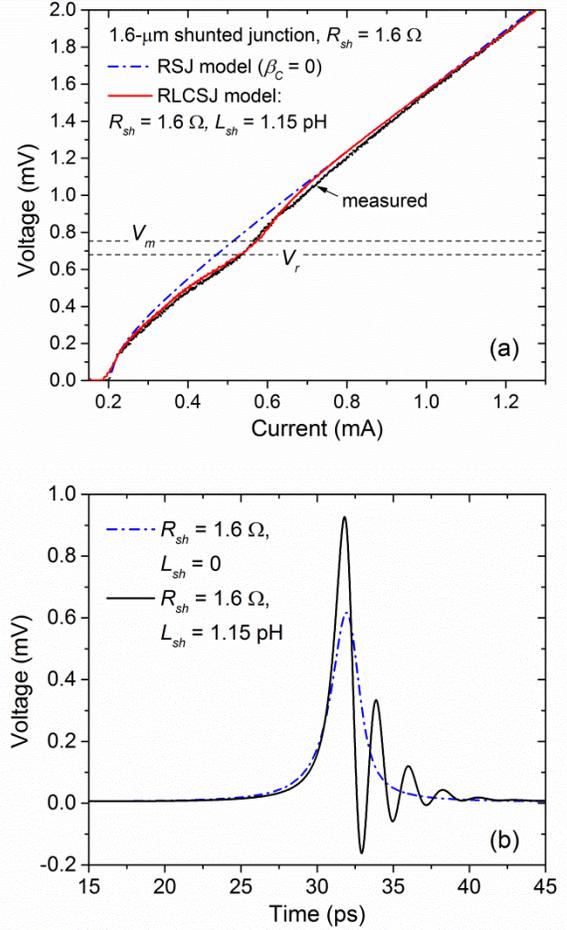

Fig. 12. (a) The measured $I$-$V$ characteristic of a 1.6-μm-diameter JJ shunted with a 1.6-Ω molybdenum-film shunt resistor and the simulated characteristics: the RSJ model ($\beta_C = \beta_L = 0$) – blue dashed-dotted curve; the RLCSJ model with the shunt inductance $L_{sh} = 1.15$ pH ($\beta_C = 0.22$, $\beta_L = 0.67$) – solid red curve. The values of the resonance voltage $V_r$ (4) and $V_m$ (6b) at these parameters are shown by the horizontal dashed lines. (b) The simulated voltage waveform (SFQ pulse shape) for a 1.6-μm-diameter JJ with shunt inductance taken into account (black solid curve) and without the shunt inductance (blue dashed-dotted curve).

The shunt inductance is usually assumed to be associated with geometrical inductance of the current loop shown in Fig. 10, [33]-[38]. In this case, the loop inductance is expected to be proportional to the loop length, $l_{loop}$, the distance from the junction edge to the I5 via along the JJ base electrode, or to the number of squares $l_{loop}/w_{M5}$

$$L_g = l_{loop} \ell + L_{via} ,\qquad (9)$$

where $\ell$ is the inductance per unit length of the microstrip formed by M5 and M6 layers, $L_{via}$ is the total contribution of all vias to the JJ and the shunt resistor, which is independent of the resistor length, and $w_{M5}$ is the wire width. Fig. 13a shows the dependence of the extracted value $L_{sh}^+$ on the loop length.

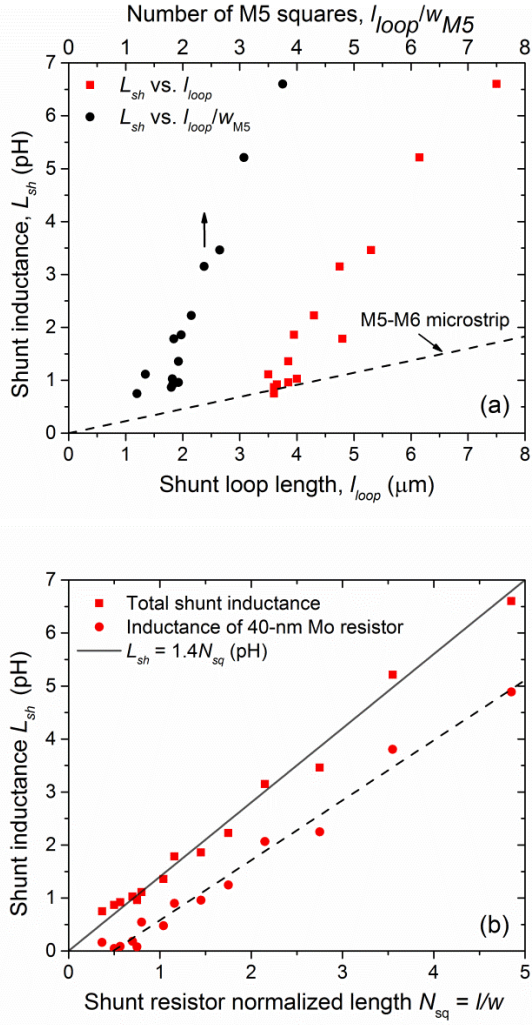

Fig. 13. Shunt inductance, $L_{sh}$ versus Mo shunt resistor design parameters. (a) $L_{sh}$ versus the shunt loop length $l_{loop}$ (bottom axis) and the number of squares $l_{loop}/w_{M5}$ in the bottom electrode (top axis); dashed line is (9) with $\ell = 0.229$ pH/µm for the M5-M6 wiring pair [39],[40]. (b) $L_{sh}$ versus the number of square of the resistor film $N_{sq} = l/w$: ■ – total inductance $L_{sh}$ (resistor + wiring); ● – contribution of the resistor film alone, $L = L_{sh} - L_g$. The solid line in (b) is the linear fit of the total inductance $L_{sh}$ versus $N_{sq}$ of the resistor film, giving $L_{sh} = 1.4 N_{sq}$ (in pH). The dashed line is the linear fit to the inductance of the Mo resistor film alone, giving inductance per square, $L_{sq} = 1.132$ pH/sq.

The expected geometrical inductance (9) is shown in Fig. 13a by the dashed line. It was calculated using 0.229 pH/µm for M5-M6 microstrips with widths $w_{M5} = w_{M6} = 2$ µm, which follows from the SQUID-based inductance measurements in [39],[40] and inductance simulations. We can see that the extracted inductance is much larger than the geometrical inductance of the loop and does not scale with loop length or the number of squares $l_{loop}/w_{M5}$. The inductance seems to extrapolate to zero at a finite length of the loop, which makes no physical sense. However, the extracted $L_{sh}$ scales linearly with the number of squares (normalized length) $l/w$ in the resistor film, as shown in Fig. 13b.

In order to get the inductive contribution of the thin normal-metal film alone, we subtracted the contribution of the M5-M6 microstrip from the extracted values of $L_{sh}$. The result is shown in Fig. 13b by solid circles. There are two potential reasons why the resultant dependence extrapolates to zero at a finite length of the resistor. The first reason is that the geometric inductance of the loop is somewhat overestimated because the length of the M6 wire $l_{loop} - l$ is shorter than the loop length, see Fig. 10. The second reason is a weak capacitive coupling between M6 wires and the resistor in a 0.35-µm overlap (surround) between the wires and the resistor between C5 vias, labeled '$sr$' in Fig. 10.

### C. 'Kinetic' Inductance of Thin-Film Shunt Resistors

It is clear from Fig. 13b that the inductance of the resistor film itself, $L = L_{sh}^+ - L_g$ is the dominant contribution to the total shunt inductance, not the inductance of superconducting Nb wiring. We suggest that this contribution is a 'kinetic' inductance of the normal-metal film forming the resistor.

In a free electron model (Drude model), the complex conductivity of a metal in an electromagnetic field of angular frequency $\omega$ is

$$\sigma(\omega) = \sigma_0/(1 + i\omega\tau), \qquad (10)$$

where $\tau$ is electron momentum relaxation (elastic scattering) time, $\sigma_0 = \frac{ne^2\tau}{m}$ is the ordinary dc conductivity, $n$ and $m$ are the electron number density and electron effective mass, respectively; see e.g., [41]. Hence, for a thin film with thickness $d$ and uniform current distribution, the imaginary part of the rf impedance $1/\sigma(\omega)$ is $\omega L = \sigma_0^{-1}\omega\tau l/(wd)$. The film inductance (Drude inductance) is simply

$$L = \left(\frac{l}{w}\right)R_{sq}\tau = N_{sq}\frac{1}{d}\frac{m}{ne^2} = N_{sq}L_{sq}, \qquad (11)$$

where $L_{sq} = m/(ne^2d)$ and $R_{sq} = 1/(\sigma_0 d)$ is the film's sheet resistance at dc.

The origin of this inductance is in electron inertia. It is an exact analog of the kinetic inductance of superconducting films and depends in the same manner on $m/n$ and $d$. This 'kinetic' inductance of normal-metal films is usually neglected because, in the typical thin films with short mean free path, $\omega\tau \ll 1$ and $\omega L_{sq} << R_{sq}$. For our 40-nm Mo film, we can estimate the elastic scattering time as $\tau = 0.57$ ps, using $L_{sq} = 1.132$ pH/sq from the fit in Fig. 13b, $R_{sq} = 2$ Ω/sq, and (11). Therefore, the condition of validity of (10), $\omega\tau \ll 1$, is satisfied at all frequencies of interest up to $\omega_L \approx 1.8 \cdot 10^{12}$ rad/s. The uniform rf current distribution is also realized because $d << \delta$, where $\delta$ is the skin depth. Thus, the $LRC$ resonance in shunted JJs gives a simple method of extracting elastic scattering time $\tau$ in various thin films by incorporating them as a JJ shunt resistor or its part.

### D. Scaling With Junction Size, $J_c$, and Sheet Resistance

The scaling with junction size is very simple and follows from (8). The required value of $\beta_c$ defines the shunt resistance. If the subgap resistance can be neglected, the shunt resistance in the first approximation is $R_{sh} \approx \frac{1}{A_J}\left(\frac{\beta_c\Phi_0}{2\pi J_c C_s}\right)^{1/2}$ and equals $N_{sq}R_{sq}$. Similarly, the shunt inductance is $L_{sh} = L_{sq}N_{sq} + L_g$. Then, in the first approximation, $\omega_r$ does not depend on the junction size and depends only on elastic scattering time and the tunnel barrier properties, and is given by

$$\omega_r^2 = \frac{1}{\tau}\left(\frac{2\pi J_c}{\beta_c \Phi_0 c_S}\right)^{1/2} - \frac{1}{\tau^2} \qquad (12)$$

This result is consistent with Fig. 4 showing that the upward bend in the *I-V* characteristics, indicating the resonance, occurs almost at the same voltage, see also Table I. In reality, $V_r$ slightly decreases because, at small resistor values (large JJ sizes), the inductance of the resistor wiring becomes comparable to the inductance of the resistor film. As a result, the quality factor increases, making $V_r$ and $V_m$ closer to each other and the resonance features more pronounced in larger JJs, as can be seen in Fig. 4.

Increasing $\beta_C$ to the typical for SFQ circuits value $\beta_C \approx 1$ by increasing the resistor length $l$, while keeping the same sheet resistance $R_{sq} = 2$ Ω/sq, increases proportionally $L_{sh}$ and decreases the resonance frequency to $V_r = 0.25$ mV, creating a rich structure in *I-V* characteristics as shown in Figs. 5 and Fig. 14. Since $\beta_C$ grows as $l^2$ and $\beta_L$ as $l$, the resonance disappears with increasing the length further.

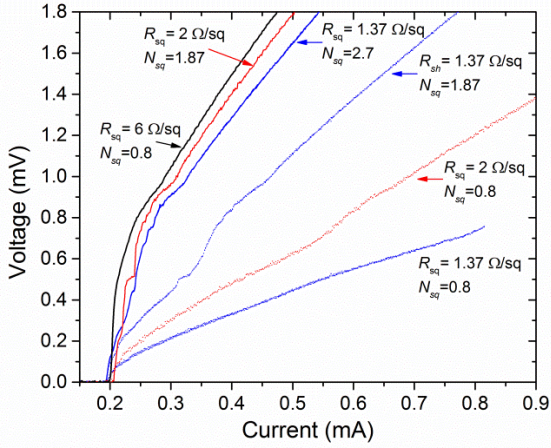

Fig. 14. *I-V* characteristics of 1.6-μm diameter junctions shunted by different shunt resistors obtained by increasing $N_{sq}$ and the sheet resistance $R_{sq}$: 1.37 Ω/sq (Mo, blue curves); 2.0 Ω/sq (Mo, red curves); 6 Ω/sq (MoN$_x$, black solid curve). The resonance features were not observed for MoN$_x$ shunts.

The most efficient way of suppressing the resonance is to increase the sheet resistance $R_{sq}$ of the resistor film by increasing electron scattering rate $\tau^{-1}$, while keeping the same film thickness and $N_{sq}$, because this increases the $\beta_C$ without increasing the $\beta_L$. We increased $R_{sq}$ by a factor of 3x, from 2 Ω/sq to 6 Ω/sq, by replacing Mo film with MoN$_x$ (nitrogen-doped molybdenum) films of the same thickness [5]. Nitrogen doping mainly increases the impurity scattering rate $\tau^{-1}$.

A comparison of the *I-V* characteristics of two junctions, one shunted by Mo resistor ($R_{sq} = 2$ Ω/sq, $\beta_C = 0.22$) and another one by MoN$_x$ resistor ($R_{sq} = 6$ Ω/sq, $\beta_C = 2$) with the absolutely identical design is shown in Fig. 14. It is clear that the resonance feature disappeared from the *I-V* curve of the MoN$_x$-shunted JJ, because MoN$_x$ provides higher shunt resistance at the same shunt inductance. In other words, the *LRC* resonance associated with the 'kinetic' inductance of the shunt resistor can be completely damped by using any resistor material having a shorter electron scattering time than the $R_{sh}C$ time constant of the junction capacitor, $R_{sh}C \geq \tau$.

In almost all shunted junctions studied, $L_{sh}$ is about a few pH. This parasitic inductance in some cases may become comparable with the signal inductances in SFQ cells. Its existence may affect timing characteristics and high-frequency dynamics of the individual cells and margins of SFQ circuits. Therefore, it may be important to include $L_{sh}$ into the junction models for SFQ circuit simulation.

## IV. JOSEPHSON PLASMA RESONANCE FREQUENCY

Knowledge of the Josephson plasma resonance frequency (voltage)

$$\omega_p = \left(\frac{2\pi I_c}{c \cdot \Phi_0}\right)^{1/2} = \left(\frac{2\pi J_c}{c_s \Phi_0}\right)^{1/2}, \qquad (13a)$$

$$V_p = (\Phi_0/2\pi)\omega_p, \qquad (13b)$$

corresponding to the internal *LC* resonance between the junction capacitance $C$ and the Josephson inductance $L_J = \Phi_0/(2\pi I_c)$, is important for circuit design in order to correctly determine the critical damping $\omega_p\tau_C = 1$ and the characteristic voltage $V_c = I_cR_s$ corresponding to the critical damping $V_c = V_p$, where $\tau_C = R_sC$.

One of the methods of measuring $\omega_p$ and the junction capacitance is to excite the Josephson plasma resonance in a large-area junction, *J2* using a small junction *J1* as a source of electromagnetic oscillations. A circuit where *J2* is coupled via resistor $R_{sh}$ was described in [42] and shown in Fig. 15. The *J2* impedance is very low at all frequencies except near the $\omega_p$ because *J2* is either shorted by its inductance or its capacitance. At $\omega_p$, the *J2* impedance becomes real, that should result in a feature in the dc *I-V* characteristic of *J1*. If both the rf and dc currents flowing through *J2* are small, an unperturbed value of (13) can be extracted. This requires the critical current of *J1*, $I_{c1}$ be much smaller that the critical current of *J2*, $I_{c2}$. The circuit was used in [15],[42],[43] for extracting $\omega_p$ and the junction specific capacitance $C_s$. Below, we will give a more detailed analysis of the circuit in order to extract $\omega_p$ and $C_s$ in high-$J_c$ junctions.

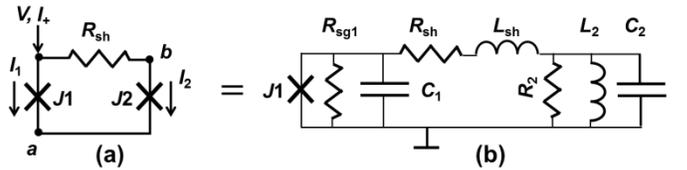

Fig. 15. A test circuit used for extracting Josephson plasma resonance frequency from *I-V* characteristics (a) and its circuit diagram (b). The dc bias current was extracted from port "*a*" making *J1* and *J2* connected in parallel at dc. Extracting the current from port "*b*" would make them connected in series as in [42],[43]. In both cases, the junctions are connected identically for the rf current generated by *J1*. In (b) *J2* is replaced by a parallel combination of junction capacitance $C_2$, Josephson inductance $L_2$, and subgap resistance $R_2$; $L_{sh}$ is the shunt inductance.

First of all, there are two possible ways of connecting *J1* and *J2* with respect to the dc bias current as shown schematically in Fig. 15a, using either port "*a*" or "*b*" as the circuit ground. The junctions are connected in parallel if the bias current is taken out of the port "*a*"; whereas they are connected in series if the current is taken out of the port "*b*" as in [42],[43]. However, the rf connection is identical in both cases. For Josephson oscillations generated by $J1$, the serially connected shunt resistor and $J2$ are in parallel to $J1$.

As a part of our standard process control monitors, we used the circuit with port "$a$" in order to reduce the current, $I_2$ flowing through $J2$ and its effect on $\omega_p$. The circuit is almost identical to the one shown in Figs. 10, 11 with the only difference being that I5 via is replaced by the large-area junction $J2$. We used two sizes of the 'active' junction $J1$, 0.7 μm and 1.0 μm, and varied the diameter of the 'passive' junction $J2$ from 2.2 μm to 6 μm.

The typical $I$-$V$ characteristics of $J1$ in a wide range of currents are shown in Fig. 16(a). The structure works as follows. At small currents $I \leq I_{c1}$, the current flows only through junction $J1$ because it is in the superconducting state. Above the critical current $I_{c1}$, a part of the total current, $I_2$ starts flowing through the shunt resistor and junction $J2$, providing a frequency-dependent damping of Josephson oscillations in $J1$. Junction $J2$ is in the superconducting state because its critical current $I_{c2} \gg I_{c1}$. Eventually, the gap voltage $V_{g1}$ is reached across $J1$, above which the $I$-$V$ characteristics becomes linear with the slope $R_n^{-1} = R_{N1}^{-1} + R_{sh}^{-1}$, shown by a dotted line in Fig. 16a, where $R_{N1}$ is the normal-state resistance of $J1$. When $I_2$ reaches $I_{c2}$, junction $J2$ switches to its gap voltage, because $J2$ is strongly underdamped. The total voltage across $J1$ jumps up by $V_{g2} = V_{g1} \approx 2.7$ mV. This is shown by the vertical arrows in Fig. 16a for the two sizes of $J2$ used. The total dc voltage across the structure after the switching is $V_{g2}^* = R_n I + V_{g2}$, as shown in Fig. 16a for $J2 = 2.2$ μm. Then, on the gap voltage branch of $J2$, the voltage across $J2$ is constant and equal $V_{g2}$ (the junction has nearly zero differential resistance). So, the $I$-$V$ characteristic remains linear with the same slope until the current through $J2$ becomes higher than the "knee" current. Above this point, the slope of the $I$-$V$ characteristic increases and becomes $V(I) = I[R_{N1}^{-1} + (R_{sh} + R_{N2})^{-1}]^{-1}$. This increase in the slope is clearly seen for $J2 = 2.2$ μm above about 0.67 mA. Upon decreasing the current, there is a big hysteresis because $J2$ remains in the resistive state and switches back into the $S$-state at a lower current as shown in Fig. 16a. So both $I_{c1}$ and $I_{c2}$, and other parameters of the test structure can be found from Fig. 16a.

A zoom in the region of voltages of interest, Fig. 16b, shows a sharp increase in dc voltage across $J1$ in the voltage range near the expected resonance voltage $V_p = \omega_p \Phi_0/(2\pi)$. Accordingly, the derivative $dV/dI$ versus $V$ displays a sharp peak at a voltage $V_{peak}$, corresponding to a resonance of a parallel type in the structure; see Fig. 16c. In [42], this peak was associated with the Josephson plasma resonance in $J2$, and its position was used to extract $C_s$ in [15],[42],[43].

However, a simple analysis shows that the resonance that really occurs in the circuit is a parallel $RLC$-type resonance discussed in III in connection with the shunt inductance. The only difference from Figs. 10,11 is that a superconducting via I5 with nearly zero inductance is replaced by the junction $J2$ with a frequency-dependent inductance, as shown in Fig. 15b.

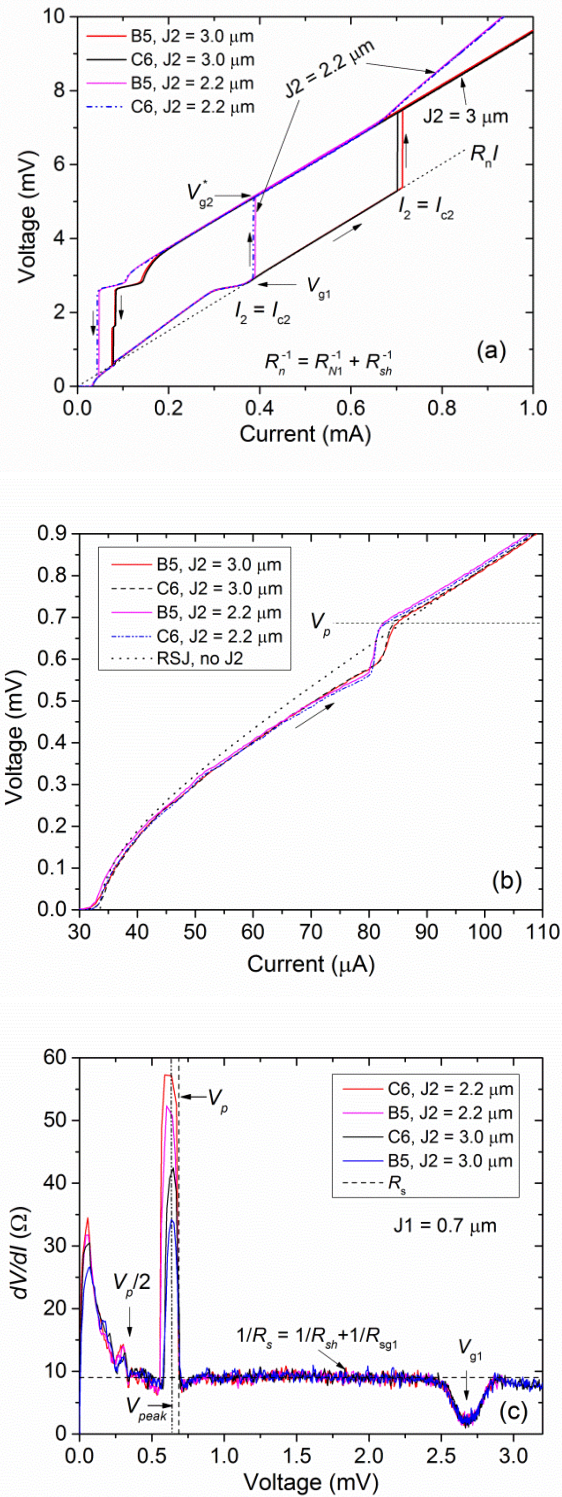

Fig. 16. (a) $I$-$V$ characteristics and $dV/dI$ for the test structures in Fig. 15 with $J1 = 0.7$ μm, $R_{sh} = 9.7$ Ω, and $J2 = 2.2$ μm and 3.0 μm at two locations, B5 and C6, on the same wafer. (b) Low-voltage part of the $I$-$V$ characteristics showing the region of the Josephson plasma resonance in $J2$; the dotted line shows the RSJ model dependence without $J2$. (c) Differential resistance $dV/dI$ vs. voltage dependences showing the peak at a voltage corresponding to a resonance of a parallel type in the structure; the curves from top to bottom correspond to the test structures in the legend.

The resonance is created by the "active" junction capacitance $C_1$ in parallel with the impedance of the shunt-$J2$ branch, $Z(\omega)$

$$Z(\omega) = R_{sh} + i\omega L_{sh} + Z_2(\omega), \qquad (14)$$

where $Z_2(\omega)$ is the equivalent impedance of $J2$,

$$Z_2(\omega) = R_2\left[\frac{(\omega/\omega_L)^2}{(1-\omega^2/\omega_p^2)^2+\omega^2/\omega_L^2} + i\frac{(\omega/\omega_L)(1-\omega^2/\omega_p^2)}{(1-\omega^2/\omega_p^2)^2+\omega^2/\omega_L^2}\right], \quad (15)$$

and $\omega_L = R_2/L_2$. The resonance frequency, $\omega_{res}$ is given by a solution of the equation

$$\omega C_1 Z(\omega)Z^*(\omega) - Im[Z(\omega)] = 0, \qquad (16)$$

which is a generalized form of (3).

At $\omega < \omega_p$, the $J2$ impedance is inductive. It adds up to $L_{sh}$, thus decreasing the circuit resonance frequency $\omega_{res}$ with respect to $\omega_r$ in (3) and increasing the $Q$-factor. The $dV/dI$ peak voltage, given by (6) for the no-$J2$ case ($L_2 = 0$), shifts down from the values given in Table I to the lower values shown in Table II and in Fig. 16c. So, essentially, the $I$-$V$ features in Figs. 16b,c are the same as those in Fig. 4, only shifted and sharpened by the presence of the $J2$ inductance.

TABLE II
PARAMETERS OF THE CIRCUIT IN FIG. 15 AT $J_c = 0.1\ mA/\mu m^2$

| J1 dia (μm) | $R_{sh}$ (Ω) | $L_{sh}$ (pH) | J2 dia (μm) | $dV/dI$ peak voltage (mV) | $V_{peak}$ Simulated (mV) | $C_s$ (fF/μm²) |
|---|---|---|---|---|---|---|
| 0.7 | 9.7 | 6.6 | 2.2 | 0.62 | 0.619 | 70 |
| 0.7 | 9.7 | 6.6 | 3.0 | 0.64 | 0.640 | 70 |
| 1.0 | 5.0 | 3.5 | 4.0 | 0.62 | 0.630 | 70 |
| 1.0 | 5.0 | 3.5 | 5.0 | 0.615 | 0.641 | 70 |

For 0.1-mA/μm² JJs, $\omega_L$ (in voltage units) is $2\pi I_c R_{sg}/\Phi_0 \sim 18$ mV. So, the $Q$-factor of the $J2$ resonator $Q_2 = \omega_L/\omega_p$ is high, $Q_2 \sim 26$. Therefore, at all frequencies (voltages) $V < V_p$, except in the immediate vicinity of $V_p$, $Im(Z) \approx \omega(L_{sh} + L_2)$. The solution of (16) can then be approximated by (3) and $V_{peak}$ value by (6) with the effective inductance $L_{sh} + L_2$.

In the limit of negligible $R_{sh}$ and $L_{sh}$, the $C_1$ capacitor is in parallel with $J2$, making the resonance at $\omega_{res} = [L_2(C_2 + C_1)]^{-1/2}$ and lowering the resonance voltage to $V_{res} = V_p/(1 + A_1/A_2)^{\frac{1}{2}}$. At intermediate values, a circuit simulator can be used to extract $C_2$ and $\omega_p$ from the peak voltage $V_{peak}$, using $L_{sh}$ values of the corresponding shunt resistors from Table I, and doing the measurements at different sizes of $J2$ and for junctions with different $J_c$ values. At voltages higher than $V_p$, the $J2$ impedance becomes capacitive and resonance with $C_1$ is no longer possible. In all the cases, the $V_p$ corresponds closely to the minimum in $dV/dI$ on the right side of the peak, as shown in Figs. 16b,c. We note that without knowledge of the value of the coupling (shunt) inductance connected to $J2$, the procedure of extracting $\omega_p$ used in [42],[43] becomes somewhat ambiguous because it involves too many fitting parameters.

The data in Fig. 16 for $J_c = 0.1$ mA/μm² serve as a self-consistency check. In III, we used $C_s = 70$ fF/μm² to extract $L_{sh}$. Now we use this $L_{sh}$ and the same $C_s$, corresponding to $V_p = 0.686$ mV, to calculate the $V_{peak}$ in a different structure. The agreement with the measurements is excellent as shown in Table II. However, in all cases, $V_{peak}$ is noticeably lower than $V_p$. Therefore, using $V_{peak}$ value instead of $V_p$ to calculate $C_s$ from (13) would noticeably overestimate the junction specific capacitance. As a consistency check we have done the measurements for different sizes of $J2$. The results are shown in Table II. As can be seen, the $dV/dI$ peak voltage very weakly depends on the $J2$ size, as expected.

In general, the plasma resonance frequency depends on the dc current through the junction $J2$ as $\omega_p^2 = (1 - I_2^2/I_{c2}^2)^{1/2}\omega_{p0}^2$, so the plasma resonance voltage shifts to a lower voltage $V_p = V_{p0}[1-(I_2/I_{c2})^2]^{1/4}$. At the resonance voltage in our structure $I_2 = V_{peak}/R_s \sim 0.1I_{c2}$, and this correction is only about 1%. We will neglect this tiny difference.

After this analysis, we turn to the higher-$J_c$ junctions, using the same circuit with exactly the same $R_{sh}$ and $L_{sh}$. With increasing $J_c$, the $I$-$V$ characteristics of the junctions and $dV/dI$ acquire additional features due to subharmonic pumping of the resonance. The typical voltage dependences of the normalized differential resistance $(dV/dI)/R_{sh}$ for the test structure in Fig. 15 with high-$J_c$ junctions are shown in Fig. 17.

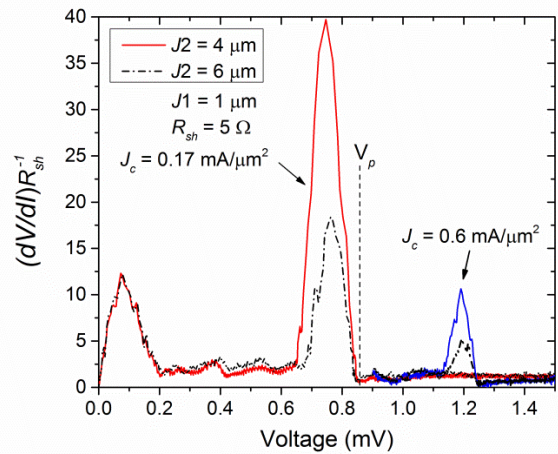

Fig. 17. Differential resistance $dV/dI$ versus $V$ dependence for the test structures in Fig. 15 with $J1 = 1.0$ μm and $J2 = 4.0$ μm (solid curves) and 6.0 μm (dash-dotted curves) at $J_c \approx 0.2$ mA/μm² (left peaks) and 0.6 mA/μm² (right peaks).

The maximum differential resistance in Figs. 16c,17 decreases with increasing the $J2$ size and/or increasing $J_c$ because the $Q$-factor decreases as a result of decreasing $R_2$ and the junction inductance $L_2$. Also, as $\omega_p$ increases with $J_c$, microwave losses in the structure also increase with frequency, reducing the $Q$-factor. At $V_p \geq \Delta/e \sim 1.35$ mV, the frequency of Josephson oscillations in the structures $2eV_p/h$ becomes larger than the absorption threshold frequency $2\Delta/h$. The dissipation strongly increases due to pair breaking, damping the plasma oscillations. This sets the maximum observable Josephson plasma resonance frequency.

The voltage of the maximum in $(dV/dI)/R_{sh}$ and the extracted values of $V_p$ are shown in Fig. 18 for the entire range of $J_c$s studied. Both quantities increase with $J_c$ but much slower than $J_c^{1/2}$ following from (13b), apparently because of increasing specific capacitance of the junctions. The $V_p$ values obtained are consistent with the data on the maximum clock frequency observed in RSFQ T-flip flops [1],[9]. The maximum value of the $V_p$ observed, $\sim 1.35$ mV is consistent with the pair-breaking criterion above.

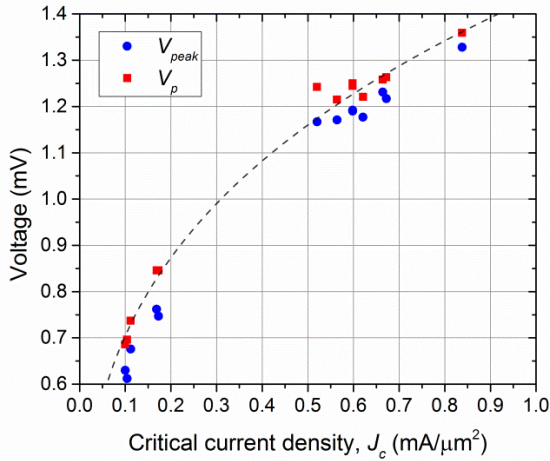

Fig. 18. The voltage $V_{peak}$ of the peak in differential resistance $dV/dI$ of the active junction $J1$ coupled to a passive junction $J2$ through the shunt resistor (blue circles) and the extracted values of the Josephson plasma resonance voltage (13b) in Nb/AlO$_x$-Al/Nb Josephson junctions with different critical current densities, $J_c$ (red squares). The dashed line is to guide the eye.

The junction specific capacitance calculated using $V_p$ and (13) is shown in Fig. 19. In the simplest model, $J_c = J_{c0}exp(-\alpha d)$ and $C_s = \varepsilon\varepsilon_0/d$, where $d$ is the tunnel barrier thickness.

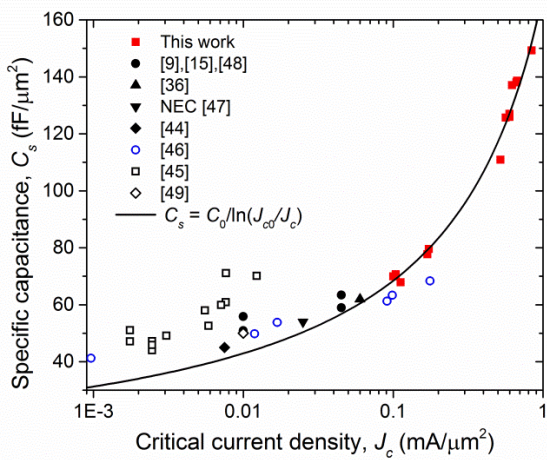

Fig. 19. The specific capacitance of Nb/AlO$_x$-Al/Nb junctions with different critical current density extracted from the $I$-$V$ characteristics of the test structures in Fig. 15. Historic data are also included. The solid line is the best fit of our data by the dependence $C_s = C_0/ln(J_{c0}/J_c)$ expected from the simplest model (17); see text.

Hence, the specific capacitance changes inversely with the logarithm of $J_c$ as

$$C_s = C_0/ln(J_{c0}/J_c) \qquad (17)$$

The solid curve in Fig. 19 shows the best fit of this dependence to our data, giving $C_0 = 265 \pm 15$ fF/$\mu$m$^2$ and $J_{c0} = 4.8 \pm 0.5$ mA/$\mu$m$^2$. For completeness we also included in Fig. 19 the literature data on the specific capacitance [9],[15],[44]-[49], obtained by using $LC$ resonance is SQUIDs and/or a zero-field resonance, Fiske step in large junctions.

## V. PARAMETER SPREADS OF HIGH-$J_c$ JUNCTIONS

The width of the distribution of the critical currents of the nominally identical junctions or, equivalently, the width of the distribution of their conductances is an important parameter characterizing the state of the fabrication technology and the maximum complexity of digital circuits [3]. As was shown in our previous work [4], these junction parameter spreads can be well described by a normal (Gaussian) distribution. The normalized (to the mean value) standard deviation of the junction normal-state resistance $\sigma_R/<R>$ is shown in Fig. 20 for junctions with sizes from 0.3 $\mu$m to 1.4 $\mu$m in diameter at three values of $J_c$ corresponding to SFQ5ee (0.1 mA/$\mu$m$^2$) and SFQ5hs (0.2 mA/$\mu$m$^2$) process node targets, and a yet higher value $J_c = 0.75$ mA/$\mu$m$^2$, corresponding to nearly the end of the $J_c$ range studied.

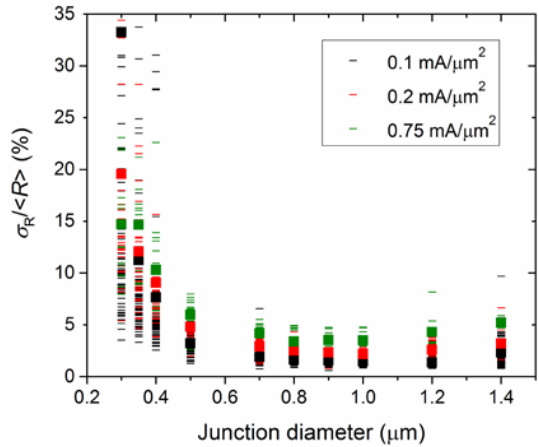

Fig. 20. Normalized standard deviation of the normal-state resistance of nominally identical junctions tested on the fabricated wafers. Each data point (dash) corresponds to the data obtained for junctions of each size located on one die. Nine die locations per 200-mm wafer were measured, 990 JJs of one given junction size per wafer. The data on 16 wafers are included: 8 wafers with $J_c = 0.1$ mA/$\mu$m$^2$ (black dashes), four wafers with $J_c = 0.2$ mA/$\mu$m$^2$ (red dashes), and four wafers with $J_c = 0.75$ mA/$\mu$m$^2$ (green dashes).

We note that the minimum junction diameter which can be printed using our deep-UV photolithography tool is 0.25 $\mu$m [4], and the minimum size allowed by the circuit design rules for our processes is 0.7 $\mu$m. It can be seen that the spreads of the junction resistances is nearly the same in junctions with $J_c = 0.1$ mA/$\mu$m$^2$ and $J_c = 0.2$ mA/$\mu$m$^2$, and only slightly higher in the highest $J_c$ case. The size-dependence is consistent with the model given in [4]. Some locations show nearly the same parameter spreads for all three current densities, including the highest one, indicating that, in some cases, the junction spreads are not yet limited by the barrier properties and can be further reduced by perfecting the fabrication process.

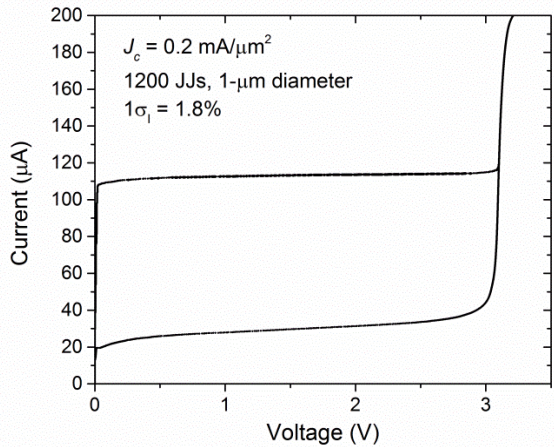

Fig. 21. *I-V* characteristics of a series array of 1200 junctions with design diameter of 1 μm and $J_c = 0.2$ mA/μm². The length of the array is about 4 mm. It characterizes the typical distribution of critical currents of the nominally identical junctions within one chip, giving a standard deviation of the critical current distribution of about 1.8%.

The results of the $I_c$ measurement at 4.2 K using an array of 1200 junctions with 1.0 μm design diameter and $J_c = 0.2$ mA/μm² are shown in Fig. 21. The results are fully consistent with the spreads obtained from the room temperature junction resistance measurements, as was shown in [4]. This once again justifies the use of room-$T$ junction resistance measurements for characterizing junction parameter spreads on the wafer scale.

A more detailed comparison of 0.1-mA/μm² and 0.5-mA/μm² junctions is given in Fig. 22 showing the aggregate distributions of room temperature resistance, $R_N$ of 0.7-μm junctions on two 200-mm wafers fabricated with these current densities. Junction resistances were normalized to the wafer mean value $< R_N >$. Locations of the junctions and their number were identical in both cases. Solid lines show fits to the normal distribution giving normalized standard deviations $\sigma_R = 2.5\%$ and 5.3%, respectively, for 0.1-mA/μm² and 0.5-mA/μm² junctions. Since these wafers were processed identically, the contribution of junction area fluctuations [4] should be identical for both current densities. Therefore, the wider statistical distributions of the higher-$J_c$ junctions can only be attributed to a higher contribution of fluctuations in AlO$_x$ barrier transparency as the barrier gets thinner.

## VI. CONCLUSION

We have studied the fabrication parameters and electrical properties of Nb/AlO$_x$-Al/Nb junctions, unshunted and with resistive shunting, in the range of Josephson critical current densities from 0.1 mA/μm² to ~ 1 mA/μm² for the use in our high-$J_c$ technology nodes. Junctions with $J_c = 0.1$ mA/μm² and 0.2 mA/μm² are highly hysteretic and require resistive shunting for applications in SFQ digital circuits. In this respect, design rules for our new 0.2-mA/μm² process SFQ5hs are identical to our 0.1-mA/μm² process SFQ5ee.

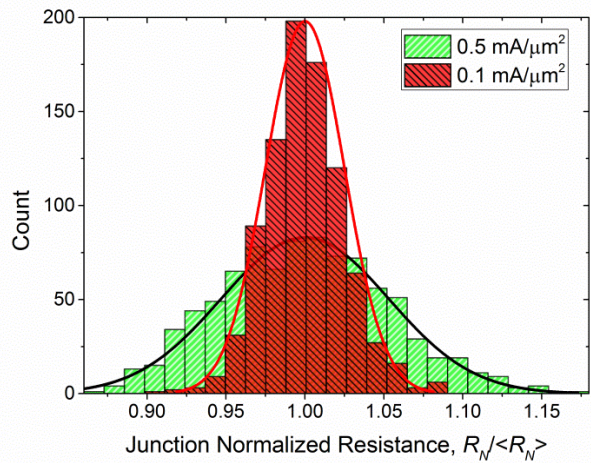

Fig. 22. Histograms of the resistance of 0.7-μm-diameter junctions comparing two 200-mm wafers, with $J_c = 0.1$ mA/μm² and $J_c = 0.5$ mA/μm². In each case 880 JJs were measured, locations for the junctions were identical in both cases. Junction resistance was normalized to the respective wafer mean value $< R_N >$. Solid lines show the fits by Gaussian distributions with $\sigma_R = 2.5\%$ and 5.3%, respectively, for 0.1-mA/μm² and 0.5-mA/μm² junctions.

We have determined the inductance associated with the resistive shunts made of thin Mo and MoN$_x$ films. We have found that the main part of the shunt inductance is the kinetic (Drude) inductance of the normal-metal film. For 40-nm Mo films this inductance is 1.13 pH/sq; it is about a factor of 3 lower for 40-nm MoN$_x$ films. The shunt inductance creates an *LRC* resonator with the junction capacitance if $\beta_L \geq \beta_C$, inducing multiple features affecting *I-V* characteristics of the shunted JJs and their switching dynamics. Although the resonance in the shunt *LRC* circuit can be easily suppressed by increasing the $\beta_C$ parameter of the shunt and using higher sheet resistance MoN$_x$ resistors, the shunt inductance remains and may significantly affect switching dynamics and timing of SFQ logic cells. We believe it should be taken into account in SFQ circuit design if accurate knowledge of the high-frequency dynamics and timing parameters is important.

Using a similar *LRC* resonance involving an additional "passive" junction *J*2 replacing the via between the shunt resistor of junction *J*1 and the bottom electrode of *J*1, we evaluated the Josephson plasma resonance frequency, junction characteristic voltage corresponding to the critical damping, and junction specific capacitance in the range of critical current densities from 0.1 mA/μm² to about 1 mA/μm². Our data are well described by the dependence $C_s = C_0/ln(J_{c0}/J_c)$ following from the simplest model, and give the fitting parameters $C_0 = 265 \pm 15$ fF/μm² and $J_{c0} = 4.8 \pm 0.5$ mA/μm². This dependence also well describes the available specific capacitance data down to $J_c \approx 1$ μA/μm², thus covering the $J_c$ range spanning about 3 orders in magnitude.

We compared the statistical distributions of the junctions of various sizes relevant to SFQ digital circuits and $J_c = 0.1, 0.2, 0.5,$ and 0.75 mA/μm². For circular junctions larger than 0.7 μm in diameter, parameter spreads of 0.1-mA/μm² and 0.2-mA/$\mu$m$^2$ junctions are essentially the same. This makes it possible to use 0.2-mA/$\mu$m$^2$ junctions in superconducting VLSI circuits. At $J_c$ of 0.5 mA/$\mu$m$^2$ and higher the distribution of junctions on-chip and on-wafer is presently a factor of two wider than for 0.1-mA/$\mu$m$^2$ junctions. This still should allow for making complex high-speed digital circuits using 0.5-mA/$\mu$m$^2$ junctions, likely with external shunts. The junction quality may further improve with process maturation.

## Acknowledgment

The authors acknowledge interesting discussions of some of the topics covered in this work with Vasili K. Semenov, Alex F. Kirichenko, Timur V. Filippov, Manjul Bhushan, and Mark B. Ketchen.